\documentclass[
    reprint,
	amsmath,
    amssymb,
    longbibliography,
	aps,
	prx,
    floatfix,
    superscriptaddress
]{revtex4-1}

\usepackage[english]{babel}
\usepackage[utf8x]{inputenc}
\usepackage[T1]{fontenc}
\usepackage{braket}
\usepackage{mathtools}
\usepackage{flushend}

\usepackage{amsmath}
\usepackage{graphicx}
\usepackage{dcolumn}
\usepackage{bm}
\usepackage{array}
\usepackage[shortlabels]{enumitem}
\newcolumntype{?}{!{\vrule width 1pt}}
\usepackage[colorinlistoftodos]{todonotes}
\usepackage{xr-hyper}
\usepackage[colorlinks=true, allcolors=black]{hyperref}

\usepackage{bibunits}
\usepackage{etoolbox}

\makeatletter
\newcommand*{\newbibstartnumber}[1]{%
  \apptocmd{\thebibliography}{%
    \global\c@NAT@ctr #1\relax
    \addtocounter{NAT@ctr}{-1}%
  }{}{}%
}
\makeatother

\newcommand\textlcsc[1]{\textsc{\MakeLowercase{#1}}}

\makeatletter
\let\cat@comma@active\@empty
\makeatother

\begin{document}
\title{Signature of Many-Body Localization of Phonons in Strongly Disordered Superlattices}
\author{Thanh Nguyen}
\affiliation{Department of Nuclear Science and Engineering, MIT, Cambridge, MA 02139, USA}
\author{Nina Andrejevic}
\affiliation{Department of Material Science and Engineering, MIT, Cambridge, MA 02139, USA}
\author{Hoi Chun Po}
\affiliation{Department of Physics, MIT, Cambridge, MA 02139, USA}
\author{Qichen Song}
\affiliation{Department of Mechanical Engineering, MIT, Cambridge, MA 02139, USA}
\author{Yoichiro Tsurimaki}
\affiliation{Department of Mechanical Engineering, MIT, Cambridge, MA 02139, USA}
\author{Nathan C. Drucker}
\affiliation{John A. Paulson School of Engineering and Applied Sciences, Harvard University, Cambridge, MA 02138, USA}
\author{Ahmet Alatas}
\affiliation{Advanced Photon Source, Argonne National Laboratory, Lemont, IL 60439, USA}
\author{Esen E. Alp}
\affiliation{Advanced Photon Source, Argonne National Laboratory, Lemont, IL 60439, USA}
\author{Bogdan M. Leu}
\affiliation{Advanced Photon Source, Argonne National Laboratory, Lemont, IL 60439, USA}
\affiliation{Department of Physics, Miami University, Oxford, OH 45056, USA}
\author{Alessandro Cunsolo}
\affiliation{Department of Physics, University of Wisconsin at Madison, Madison, WI 53706, USA}
\author{Yong Q. Cai}
\affiliation{National Synchrotron Light Source II, Brookhaven National Laboratory, Upton, NY 11973, USA}
\author{Lijun Wu}
\affiliation{Condensed Matter Physics and Material Science Department, Brookhaven National Laboratory, Upton, NY 11973, USA}
\author{Joseph A. Garlow}
\affiliation{Condensed Matter Physics and Material Science Department, Brookhaven National Laboratory, Upton, NY 11973, USA}
\author{Yimei Zhu}
\affiliation{Condensed Matter Physics and Material Science Department, Brookhaven National Laboratory, Upton, NY 11973, USA}
\author{Hong Lu}
\affiliation{College of Engineering and Applied Sciences, Nanjing University, Nanjing, China}
\author{Arthur C. Gossard}
\affiliation{Materials Department, University of California, Santa Barbara, Santa Barbara, CA 93106, USA}
\author{Alexander A. Puretzky}
\affiliation{Center for Nanophase Materials Sciences, Oak Ridge National Laboratory, Oak Ridge, TN 37831, USA}
\author{David B. Geohegan}
\affiliation{Center for Nanophase Materials Sciences, Oak Ridge National Laboratory, Oak Ridge, TN 37831, USA}
\author{Shengxi Huang}
\thanks{Corresponding authors.\\ \href{mailto:mingda@mit.edu}{mingda@mit.edu}, \href{mailto:sjh5899@psu.edu}{sjh5899@psu.edu} \vspace{0.5cm}}
\affiliation{Department of Electrical Engineering, The Pennsylvania State University, University Park, PA 16802, USA}
\author{Mingda Li}
\thanks{Corresponding authors.\\ \href{mailto:mingda@mit.edu}{mingda@mit.edu}, \href{mailto:sjh5899@psu.edu}{sjh5899@psu.edu} \vspace{0.5cm}}
\affiliation{Department of Nuclear Science and Engineering, MIT, Cambridge, MA 02139, USA}
\date{\today}

\begin{abstract}
\noindent \textbf{Abstract} \\
Many-body localization (MBL) has attracted significant attention due to its immunity to thermalization, role in logarithmic entanglement entropy growth, and opportunities to reach exotic quantum orders. However, experimental realization of MBL in solid-state systems has remained challenging. Here we report evidence of a possible phonon MBL phase in disordered GaAs/AlAs superlattices. Through grazing-incidence inelastic X-ray scattering, we observe a strong deviation of the phonon population from equilibrium in samples doped with ErAs nanodots at low temperature, signaling a departure from thermalization. This behavior occurs within finite phonon energy and wavevector windows, suggesting a localization-thermalization crossover. We support our observation by proposing a theoretical model for the effective phonon Hamiltonian in disordered superlattices, and showing that it can be mapped exactly to a disordered 1D Bose-Hubbard model with a known MBL phase. Our work provides momentum-resolved experimental evidence of phonon localization, extending the scope of MBL to disordered solid-state systems. \\

\noindent \textit{\textbf{Keywords}: many-body localization, x-ray scattering, phonon, superlattice, Bose-Hubbard model}

\end{abstract}

\maketitle

%%%%%%%%%%%%%%%%%%%%%%%%%%%%%%%%%%%%%%%%%%%%%%%%%%%%%%%%%%%%%%%%%%%%%%%%%%%%%%%%%%%%%%%%%%%%%%%%%%%%%%%%%%%%%%%%%%%%%%%%%%%%%%%%%%%%%%%%%%%%%%%%%%%%%%%%%%%%%%%%%%%%%%%%%%%%%%%%%%%%%%%%%%%%%%%%%%%%%%%%%%%%%%%%%%%%%%%%%%%%%%%%%%%%%%%%%%%%%%%%%%%%%%%%%%%%%%%%%%%%%%%%%%%%%%%%%%%%%%%%%%%%%%%%%%%%%%%%%%%%%%%%%%%%%%%%%%%%%%%%%%%%%%%%

\noindent Many-body localized (MBL) states subject to strong disorder and strong interactions provide a platform for highly unusual phenomenology \cite{basko2006,pal2010,nandkishore2015,abanin2019,khemani2020}. Notable examples include the absence of thermalization upon evolution under intrinsic dynamics for parametrically long time scales, quantum phase transitions at nonzero temperatures, and novel dynamical phases of matter. In particular, MBL can protect exotic quantum orders through localization of excitations that would otherwise result in trivialization of the phases \cite{huse2013,chandran2014,pekker2014}. Hence, these systems have recently seen heightened interest by serving as a possible pathway towards quantum information processing in systems driven far from equilibrium. MBL has been the subject of exciting studies in well-controlled artificial quantum systems such as ultracold atoms \cite{kondov2015,schreiber2015,choi2016,lukin2019,rispoli2019}, trapped ions \cite{smith2016}, interacting spin chains \cite{wei2018}, and superconducting circuits \cite{roushan2017,xu2018}. In contrast to such isolated systems, the unavoidable environmental couplings in a solid-state system smear the sharp features of MBL \cite{nandkishore2017} and thus often preclude further experimental explorations.

In parallel with MBL, the importance of phonon localization can hardly be overlooked. It is well known that interference between multiple wave scattering paths in systems with even arbitrarily weak disorder can result in Anderson localization depending on the system dimensionality \cite{anderson1958,abrahams1979,lee1985}. As a fundamental wave phenomenon, Anderson localization has been theoretically studied in a wide variety of systems \cite{cutler1969,thouless1974,sutherland1986,john1987,weaver1990,sanchez-palencia2007}, including phonons \cite{john1983,kirkpatrick1985,sheng1994}. However, unlike photons \cite{albada1985,wolf1985,wiersma1997,schwartz2007,segev2013} and classical waves \cite{hu2008} where coherence can be readily maintained, direct measurements of phonon localization had eluded experiments until recently. The observation of coherent phonons in semiconducting superlattice (SL) systems enabled the exploration of phonon-related phenomena that relied on such coherence \cite{luckyanova2012}. Numerical calculations \cite{mendoza2016,hu2018,juntunen2019} alongside experimental evidence in heat conduction \cite{luckyanova2018} established localization as a fundamental mode of phonon transport. 

\begin{figure*}[t!]
	\centering
	\includegraphics[width=\linewidth]{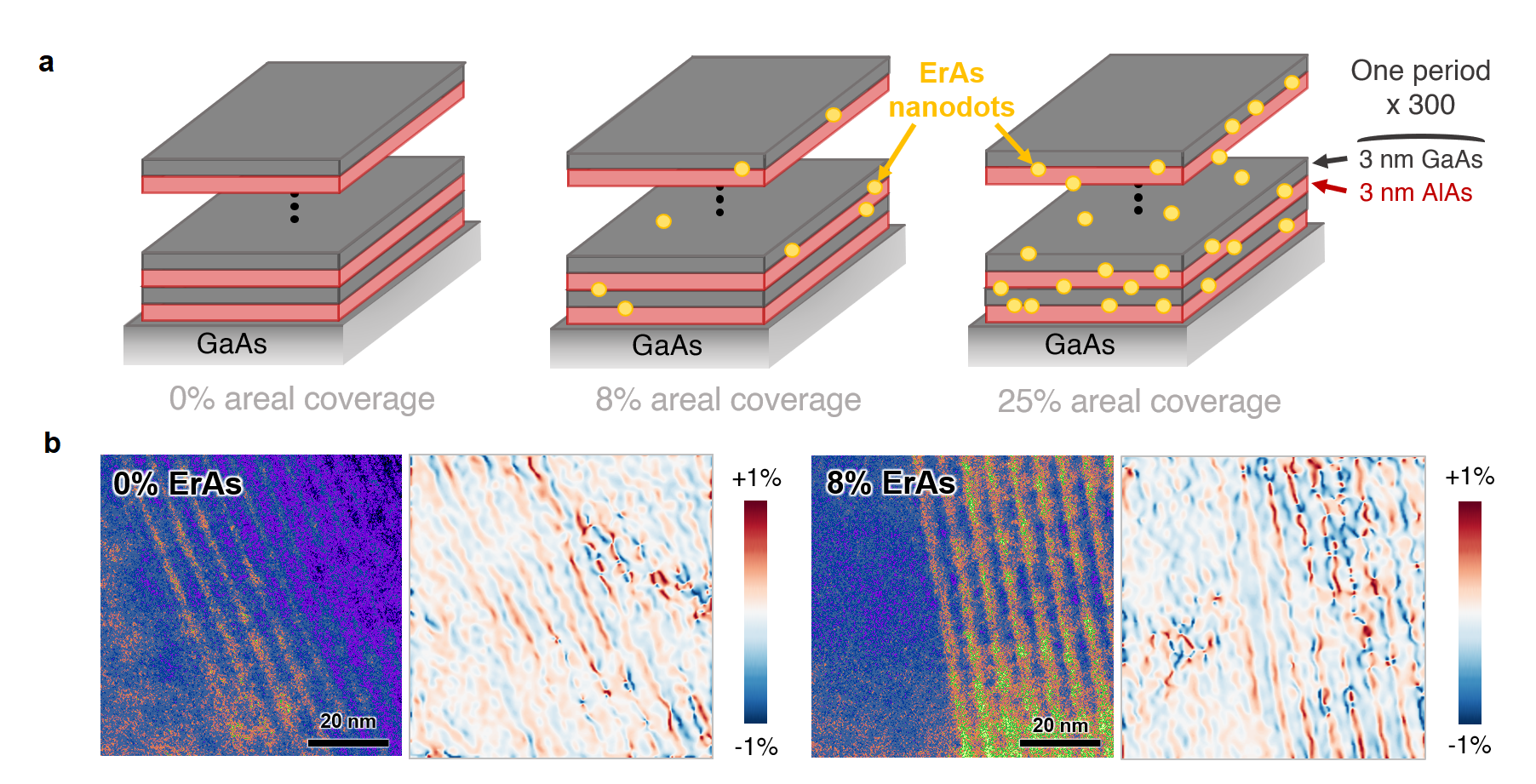}
	\caption{\textbf{GaAs/AlAs SL samples.} \textbf{a}, Illustration of the pristine Ref (0\%) and disordered GaAs/AlAs SL samples with 3 nm-diameter ErAs nanodots (8\% and 25\% areal coverage). The SLs consist of 300 periods of 3 nm-thick GaAs and 3 nm-thick AlAs layers grown on a GaAs substrate. \textbf{b}, Transmission electron micrograph and related strain mapping of the Ref (left two images) and 8\% ErAs disordered SLs (right two images).}
	\label{fig:1}
\end{figure*}

Nevertheless, open questions remain. Theoretical predictions suggest that phonon localization would transpire within a finite energy interval characterized by multiple mobility edges separating propagating from localized states \cite{kirkpatrick1985,mendoza2016}, but momentum-resolved phonon localization is yet to be demonstrated. The strong interactions and strong disorders in phononic SL systems further calls for an MBL description of phonon localization. The elucidation of phonon localization through the lens of MBL would have major implications for connecting the MBL phenomenology to a mesoscopic solid-state system and can shed light on further avenues towards thermoelectric energy conversion and thermal management in microelectronics \cite{maldovan2013}.

In this work, we experimentally demonstrate a strong departure from phonon thermalization in disordered GaAs/AlAs SLs embedded with ErAs nanodots using grazing-incidence inelastic x-ray scattering (GI-IXS), which reveals a possible manifestation of MBL. Such measurements are technically challenging as they require resolving energy shifts on the order of meV with $\sim$10 keV x-rays scattered off a $\mu$m-thick thin film \cite{serrano2011}. Comparison of the phonon Stokes and anti-Stokes signals reveals persistent non-equilibrium conditions of the phonon population when disordered ErAs nanodots are introduced. This is suggestive of an MBL phase which is characterized by an effective temperature that deviates from its thermal value in a canonical ensemble \cite{rossini2009,canovi2012,lenarcic2018,lenarcic2020}. This anomalous behavior occurs systematically within a range of phonon energies and wavevectors at low temperature, revealing a crossover behavior of MBL. We support our experimental results through an exact theoretical mapping between an effective phonon Hamiltonian for disordered SLs and the 1D Bose-Hubbard model with a known MBL phase. Our work presents strongly disordered SL systems as a promising setting to further explore MBL phenomena in solid-state systems.\\

\textit{Sample characterization.} The samples in the GI-IXS experiments comprise three sets of 300-period GaAs/AlAs SLs epitaxially grown on a GaAs (001) substrate. They possess a superperiodicity of 6 nm consisting of a 3 nm GaAs layer followed by a 3 nm AlAs layer. Two disordered SLs are embedded with randomly-distributed 3 nm-diameter ErAs nanodots: one with an areal density of 8\% and the other, 25\% (Fig. \hyperref[fig:1]{1a}). The average ErAs interparticle distance is 9.5 nm and 5.5 nm, respectively. The third, pristine SL with 0\% nanodot density is kept as a reference (hereafter referred to as "Ref"). All samples were grown by molecular beam epitaxy (MBE) in a Veeco Gen III MBE system. The sub-monolayer ErAs deposition was inserted at the SL interfaces, and the formation of nanodots are caused by the self-assembly of deposited atoms. The size of the ErAs nanodots and the inter-nanodot distances are comparable to the SL superperiodicity, indicating a strong effect of ErAs disorder, in contrast to point-like impurities. Transmission electron microscope-based strain mapping (Fig. \hyperref[fig:1]{1b}) shows a low and uniform strain distribution across the SLs, indicating high-quality growth of disordered-SL at similar strain level of Ref-SL. Further characterizations of these samples can be found in \cite{luckyanova2018}. 

\begin{figure}[ht!]
	\centering
	\includegraphics[width=\linewidth]{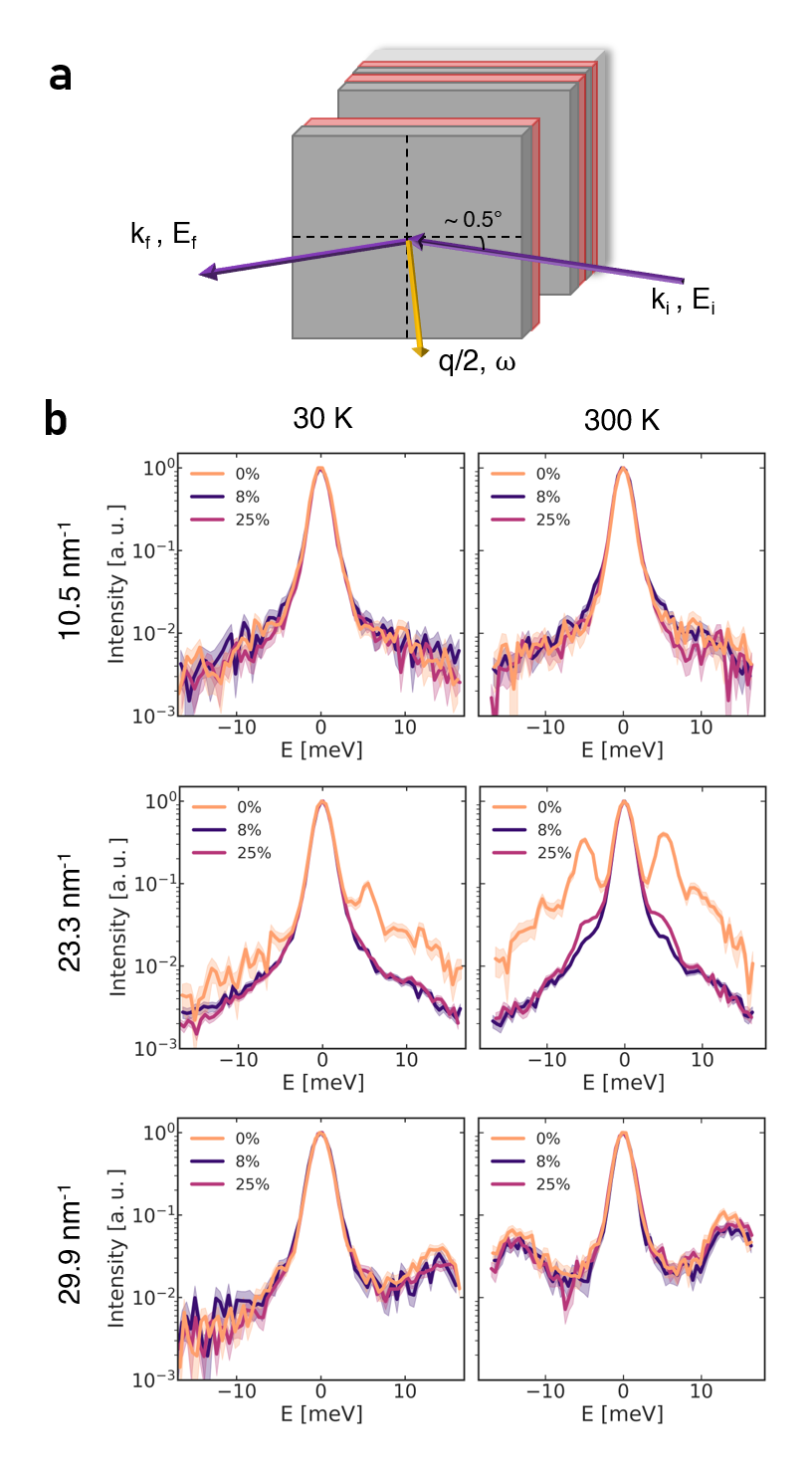}
	\caption{\textbf{GI-IXS of the GaAs/AlAs SLs.} \textbf{a}, Scattering geometry of the synchrotron-based GI-IXS on the SLs. \textbf{b}, Comparison of constant-$q$ scans of samples with 0\%, 8\% and 25\% ErAs nanodot density at representative $q$ values at 30 K (left column) and 300 K (right column). At medium-$q$ regime near $q = 23.3$ nm$^{-1}$, there is a noticeable departure of the phonon spectra between the disordered SLs and the Ref-SL. Most importantly, the intensity ratio between the Stokes and the anti-Stokes peaks show anomalous behavior in disordered samples at low temperature. The intensity is normalized and shaded pale regions indicate one standard deviation.}
	\label{fig:2}
\end{figure}

Non-specular GI-IXS experiments were carried out at beamline 3-ID-C in Advanced Photon Source at Argonne National Lab \cite{Sinn2001,Alatas2011,Toellner2011,Sinn2001b}. A schematic is illustrated in Fig. \hyperref[fig:2]{2a}. With an incident x-ray energy of $E_i=$ 21.7 keV, a grazing incident angle of $\theta_i=0.5^{\circ}$ was chosen such that the corresponding probing depths $\lambda_a(\text{GaAs})=0.46$  $\mu$m and $\lambda_a(\text{AlAs})=0.82$ $\mu$m are comparable to the 1.8 $\mu$m total SL thickness. This indicates a dominant scattering contribution from the SL rather than from the substrate. Given the large $E_i$ (corresponding to a wavevector $k_i = 110$ nm$^{-1}$) and small $\theta_i$, the x-ray wavevector transfer $q \equiv k_i-k_f$ is largely in the out-of-plane direction of the SL in the measured range up to 29.9 nm$^{-1}$, indicating quasi-1D phonon measurements (calculated phonon dispersions and Raman measurements in the Supporting Information). Constant-$q$ scans of the energy transfer ($E \equiv E_i-E_f$) were performed for all samples at eight different $q$ values, of which three representative ones at low, medium and high values are shown in Fig. \hyperref[fig:2]{2b} (other values in Fig. S1). These measurements were performed at an optimized low temperature of 30 K to balance between the long phonon coherence length and the high phonon population \cite{mendoza2016}, and at room temperature. 

At low-$q$ values ($q=10.5$ nm$^{-1}$ in Fig. \hyperref[fig:2]{2b}), the spectra of the Ref SL and disordered SLs behave similarly by nearly overlapping each other. The low phonon and high elastic peak intensities make it difficult to quantitatively extract the phonon dispersion in this SL system. As one accesses medium-$q$ values ($q=23.3$ nm$^{-1}$ in Fig. \hyperref[fig:2]{2b}), the behavior of the anti-Stokes ($E<0$) and Stokes ($E>0$) peaks become highly unusual. Thermodynamic equilibrium dictates that the Stokes/anti-Stokes amplitude ratio satisfies detailed balance where at high temperatures, this ratio approaches unity, and at low temperatures, the Stokes peak intensity dominates over that of the anti-Stokes signal. All samples, irrespective of ErAs content, follow this tendency at low-$q$ and high-$q$ regimes for both temperatures. However, two key discrepancies are discernible in the medium-$q$ range. First, whereas the Ref SL sample shows significantly larger Stokes peak intensity at low temperature, the disordered SL samples have surprisingly comparable intensities for both anti-Stokes and Stokes peaks at low temperature, highlighting a departure from thermal equilibrium. An attribution to the exciton resonance effect \cite{goldstein2016} is easily  excluded due to the distinct energy scale and temperature profile, and all measured SLs are electrically insulating with no exciton effects. Second, the presence of disorder drastically decreases phonon intensities relative to the elastic peak, indicating a suppressed phonon population \cite{Sandro2018}. This behavior is most prominent at $q=23.3$ nm$^{-1}$ but can also be found in nearby $q$ values. This would point towards delimiting a finite regime of wavevectors where the phonons do not properly thermalize. Moreover, the phonon spectra at $q=23.3$ nm$^{-1}$ and low temperature has a broad linewidth indicating a high relaxation rate, which further suggests a strong phonon interaction that may facilitate the MBL formation. At high-$q$ values ($q=29.9$ nm$^{-1}$ in Fig. \hyperref[fig:2]{2b}), the spectra between the Ref SL and disordered SLs largely overlap again. A small energy shift between the Ref SL and disordered SLs is observed, which may be attributed to the effect of disorder on the phonon energy. All of these conclusions so far can be drawn from raw data prior to further data analysis.\\

\begin{figure}[h!]
	\centering
	\includegraphics[width=\linewidth]{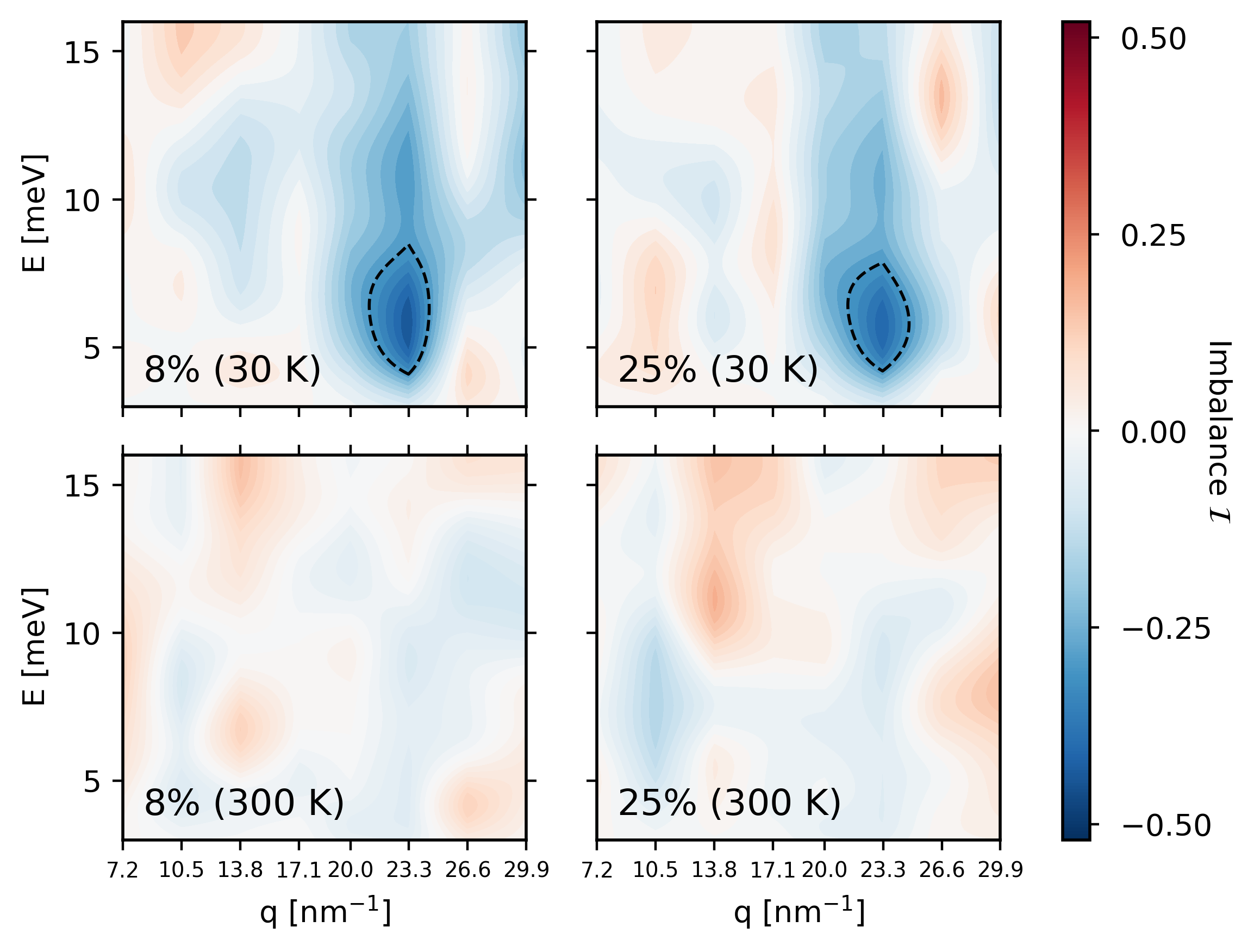}
	\caption{\textbf{Phonon population imbalance in energy and wavevector parameter space.} Phonon population imbalance $\mathcal{I}$ for ErAs-disordered SLs (8\%, left column, and 25\%, right column) as defined in Eq. \eqref{eq:pop_imbalance}. The imbalance is extracted using the data in Fig. \ref{fig:2}. The high imbalance $\mathcal{I}$ is only observed at low temperature (top row) but not at room temperature (bottom row), while the two disordered SLs show large imbalance up to $|\mathcal{I}| \sim 0.5$ in similar energy window near $E = 6$ meV and wavevector window  $q = 23$ nm$^{-1}$. The dashed lines denote the contour for $|\mathcal{I}|=0.3$.}
	\label{fig:3}
\end{figure}

\textit{Observation of localization-thermalization crossover.} To reinforce the aforementioned claims, we quantify the difference in behavior of the Stokes and the anti-Stokes peak intensities between the Ref and disordered samples. The ratio $R=$ I(Stokes)/I(anti-Stokes) for each sample is determined from the GI-IXS spectra as a function of the phonon energy for each constant-$q$ scan. The calculation of $R$ is performed without removing the elastic peak to preclude any undesirable artifacts that could arise from elastic-peak subtraction (further analysis in the Supporting Information). The phonons in the Ref SL follow equilibrium statistics based on previous evidence \cite{luckyanova2012}. The data are presented to display energy ranges outside those which may be significantly impinged by the elastic peak. To characterize the phonon population imbalance of the disordered samples, we define an imbalance parameter, similar to other studies \cite{levi2016,luschen2017}, as 
\begin{equation}
    \mathcal{I} = \frac{R_{\text{Disorder}}-R_{\text{Ref}}}{R_{\text{Disorder}}+R_{\text{Ref}}},
    \label{eq:pop_imbalance}
\end{equation}
\noindent where $R_{\text{Ref}}$ refers to the Ref SL and $R_{\text{Disorder}}$ refers to disordered SLs with either 8\% or 25\% disorder density by area coverage. Plots of this parameter in energy-wavevector space for both disordered samples are shown in Fig. \ref{fig:3}. Both display $|\mathcal{I}| < 0.15$ at all measured energies and wavevectors at room temperature, where localization is not expected due to strong temperature effects towards equilibrium. Additionally, the GaAs substrate does not have an effect on the measured value of $\mathcal{I}$ (Supporting Information).

As noted from the GI-IXS spectra, the largest imbalance in the phonon population occurs near $q=$ 23.3 nm$^{-1}$ (line cut with error bars in Fig. S2), reaching absolute values up to $|\mathcal{I}| \sim$ 0.5. This value is comparable to analogous number imbalance measurements within the MBL phase in ultracold atom experiments \cite{schreiber2015,choi2016}. The SL with lower disorder density (8\%) appears to experience slightly larger imbalance in a more prominent region in energy-wavevector phase space, but the general trend between the two disordered samples is very similar. Moreover, in addition to the population imbalance taking place within a medium-$q$ window, the behavior also occurs in a finite energy interval between $\sim$4-13 meV. Fig. \ref{fig:3} shows distinct regimes in the energy-wavevector phase space with large (dark blue) or weak (light colors) imbalance, which may further serve as a boundary between MBL and thermal phases. As the phononic SL system on a substrate is more akin to an open quantum system, one expects a crossover behavior instead of a sharp transition between the thermal and the MBL phases as in a closed quantum system. Nevertheless, the suppression of phonon population at medium-$q$ regime is indicative of a possible reduction in the effective dimensionality of the site-Hilbert space, which could be a manifestation of a strong local repulsion (Fig. \ref{fig:2}b).\\

\textit{Effective phonon action in disordered SLs.} To gain further insight on the phonon behavior, we develop a minimal model for the phonon transport in the disordered quasi-1D SL system using a tight-binding description. The suitability of the tight-binding approach can be justified as the computed minimum localization length of $\sim$20 nm in disordered SL samples \cite{luckyanova2018} is comparable to the characteristic length scale of the SL system with a superperiodicity of 6 nm. The tight-binding phonon Hamiltonian of the Ref SL can be written as
\begin{equation}
    H_0 = \sum_i \varepsilon_in_i + \sum_i t_{i}\left(a_{i+1}^\dagger a_i+a_i^\dagger a_{i+1}\right),
    \label{eq:freeTB}
\end{equation}
\noindent where $\varepsilon_i$ is the on-site energy of the \textit{i}\textsuperscript{th} SL layer, $n_i=a_i^\dagger a_i$ is the local phonon population, and $t_{i}$ is the hopping amplitude between the neighboring SL layers. We further introduce the disorder-field Hamiltonian $H_d$ and the phonon-disorder interaction Hamiltonian $H_I$ as
\begin{equation}
\begin{split}
    H_{d} &=\frac{1}{2}\sum_i\varepsilon_{d, i}\xi_i^2 \\
    H_I &= \sum_i g_{i}a_i^\dagger a_i\xi_i, 
    \label{eq:disorder}
\end{split}
\end{equation}
\noindent where $\varepsilon_{d,i}>0$ is the on-site disorder energy of the \textit{i}\textsuperscript{th} SL layer, $\xi_i$ is the disorder field, and $g_{i}$ is the local phonon-disorder coupling strength. The on-site disorder energy $\varepsilon_{d,i}$ is appropriate due to the mass difference by the ErAs nanodots. The phonon-disorder coupling is proportional to the phonon population $n_i$, which is also reasonable. 

For a comprehensive impurity average calculation, either replica theory \cite{Mezard1987,Dotsenko2001}, Keldysh formalism \cite{kamenev2011}, or supersymmetry method \cite{Efetov1983} should be implemented. Here for a qualitative estimate without loss of generality, we define the effective phonon action $S_{\text{eff}}$ by integrating over the disorder degrees of freedom: 
\begin{equation}
    e^{-S_{\text{eff}}}=e^{-S_0}\times\int D\xi\ e^{-S_{d}[\xi]-S_I[\xi]},
    \label{eq:effaction}
\end{equation}
\noindent where $D\xi=\prod_i d\xi_i$ is the functional measure. Such treatment makes the disorder field resemble annealed disorder. By carrying out an explicit calculation of Eq. \eqref{eq:effaction}, and further defining that $J = -t_i, U_i = g_{i}^2/\varepsilon_{d,i},
    \mu_i =\varepsilon_i + g_{i}^2/(2\varepsilon_{d,i})$, we obtain
the 1D Bose-Hubbard model

\begin{align}
\begin{split}
    H_{\text{BH}} = & \sum_i \mu_in_i -J\sum_i\left(a_{i+1}^\dagger a_i+a_i^\dagger a_{i+1}\right) \\
    &+ \frac{1}{2}\sum_i U_i\left(n_i^2 - n_i\right).
    \label{eq:bose-hubbard}
\end{split}
\end{align}
This model has been extensively studied and shown to exhibit an MBL phase \cite{michal2014,michal2016}. Furthermore, the MBL in this model is characterized by an \textit{inverted} mobility edge in which higher energy states are localized \cite{sierant2017,sierant2018}, similar to what is observed in our GI-IXS experiment.

The on-site repulsion $U_i$ in Eq. \eqref{eq:bose-hubbard} is an important parameter controlling the model behavior, and can be estimated as (Supporting Information)
\begin{equation}
    U_i \propto \frac{R^2}{N n_B(\varepsilon_{d,i})} \sqrt{\frac{m_{\text{Er}}}{m_{\text{eff}}}},
    \label{eq:Ui}
\end{equation}
 
\noindent where $R$ is the nanodot radius, $n_B$ is the Bose-Einstein distribution, $m_{\text{Er}}$ is the mass of Er atom from disorders and $m_{\text{eff}}$ is the effective mass from Al and Ga atoms from Ref SLs, and $N$ denotes the system size. A number of experimental features can be reconciled with Eq. \eqref{eq:Ui}. First, the disorder concentration $\eta$ is cancelled out and absent in Eq. \eqref{eq:Ui}, which is consistent with the similar behavior observed in our 8\% and 25\% samples in the parameter regime with large phonon population imbalance. Second, the large mass differences between Er atoms and the Ga and Al atoms tend to increase $U_i$ and favor MBL. Third, the large disorder radius $R$ with an  $R^2$-dependence also facilitates a large $U_i$. Lastly, a lower phonon population inside the disorder $n_B(\epsilon_{d,i})$ can generate a larger $U_i$, which agrees with the observation that at medium-$q$ phonon population is suppressed while imbalance $\mathcal{I}$ is high. The possibility to exactly map the strongly disordered quasi-1D phonon transport into the Bose-Hubbard model, as well as the guideline to increase on-site repulsion using Eq. \eqref{eq:Ui}, hints at a new avenue to seek MBL in disordered SL solid-state systems.\\

\textit{Discussion.} In our SL system, the natural choice of system size is the total period of SL $N=300$. However, coherent phonon behavior can be observed in an $N=16$ SL \cite{luckyanova2012}, within which the inter-SL transport can be "coarse-grained" as one effective site. In accordance with the long phonon coherence and long phonon mean-free-path $l_{\text{MFP}}$ in an SL with total thickness $L$ \cite{mendoza2016}, the effective system size for a $N=300$ period SL gives $N_{\text{eff}} = N l_{\text{MFP}}/L < 20$ , which serves the role of $N$ in Eq. \eqref{eq:Ui} and further increases the on-site repulsion. In addition, the observed imbalance $\mathcal{I}$ takes on an extreme value near $E=5$ meV, with $\mathcal{I}=-0.5$, while it diminishes to $\mathcal{I}=-0.15$ near $E=15$ meV, indicating the existence of two mobility edges in energy. Although the 1D Bose-Hubbard model only has one inverted mobility edge, the additional mobility edge can be intuitively understood as resulting from high-energy extended modes which can be added to Eq. \eqref{eq:freeTB} as longer-range hopping terms. This behavior qualitatively agrees with the self-consistent diagrammatic theory of phonon localization in hard-sphere scatterers \cite{kirkpatrick1985}, and semi-quantitatively with the non-equilibrium Green's function calculations in this SL system showing the localization regime at around 6 to 10 meV \cite{mendoza2016}. On the other hand, the localization window in wavevector space has not been predicted since theories either integrate over momentum degrees of freedom or set the momentum to zero for simplicity \cite{kirkpatrick1985,john1983}. In our 6 nm superperiodic SL, $q \sim 20$ nm$^{-1}$ corresponds to a $\sim$40-fold extended Brillouin zone. From the computed SL phonon dispersion \cite{luckyanova2018}, this enters into the phonon energy at $E \sim 10$ meV, above the lower-energy delocalized regime. Consequently, the localization window in wavevector space may naturally be linked to the localization window in energy space.\\

\textit{Conclusion.} In summary, we provide momentum-resolved evidence of signatures of many-body localization in disordered semiconducting phononic SLs, through state-of-the-art low temperature GI-IXS measurements. The disordered SL platform contains the essential elements that can lead to the realization of MBL: a reduced dimensionality which limits the pathways to drive the system towards thermal equilibrium, while the nanodot-type disorder with size comparable  to the SL spacing can provide strong on-site interaction and strong disorder effects. By investigating the quasiparticle population imbalance, we uncover that phonons in the disordered SL samples fail to thermalize within finite energy and wavevector windows. This likening of the experimental observations to the MBL phenomenology is further supported by theoretical calculations showing the equivalence between the quasi-1D tight-binding phonon transport under strong disorder and the 1D Bose-Hubbard model with a known MBL phase. In particular, our experimental findings qualitatively agree with the criteria to reach high on-site repulsion that favors an MBL phase. Our study opens up new opportunities in two directions: as a new experimental solid-state platform for realizing possible MBL phenomena, and as a novel phonon phase far away from equilibrium.

\section*{Supporting Information}
The Supporting Information is available free of charge at http://pubs.acs.org.\\

Parameter estimation of the strong Hubbard interaction, data analysis of GI-IXS measurements, phonon dispersion of GaAs/AlAs/ErAs superlattices, effect of GaAs substrate, and Raman scattering.

\section*{Acknowledgments}
The authors thank R. Nandkishore and P. Cappellaro for helpful discussions. T.N., N.A., N.C.D. and M.L. acknowledge the support from U.S. Department of Energy (DOE), Office of Science, Basic Energy Sciences (BES), award No. DE-SC0020148. N.A. acknowledges the support of the National Science Foundation Graduate Research Fellowship Program under Grant No. 1122374. H.C.P. is supported by a Pappalardo Fellowship at MIT and a Croucher Foundation Fellowship. Work of Q.S. and Y.T. was supported by Solid State Solar-Thermal Energy Conversion Center (S3TEC), an Energy Frontier Research Center funded by the U.S. Department of Energy, Office of Science, Basic Energy Sciences, award No. DE-SC0001299 (prior to January 2019). L.W., J.G., and Y. Z. acknowledge  the support from DOE/BES, the Materials Science and Engineering Divisions,  under Contract No. DE-SC0012704. This research used resources of the Advanced Photon Source, a U.S. DOE Office of Science User Facility operated for the DOE Office of Science by Argonne National Laboratory under Contract No. DE-AC02-06CH11357. Raman measurements were conducted at the Center for Nanophase Materials Sciences (CNMS), which is a DOE Office of Science User Facility.

\bibliographystyle{apsrev4-1}
\bibliography{references}

\clearpage
\onecolumngrid

\setcounter{equation}{0}
\setcounter{figure}{0}
\setcounter{table}{0}
\renewcommand{\theequation}{S\arabic{equation}}
\renewcommand{\thefigure}{S\arabic{figure}}
\renewcommand{\thetable}{S\arabic{table}}

\newcounter{SIfig}
\renewcommand{\theSIfig}{S\arabic{SIfig}}

\section*{Signature of Many-Body Localization of Phonons in Strongly Disordered Superlattices: Supplemental Materials}

\subsection{Parameter estimation of the strong Hubbard interaction}
\noindent To evaluate the magnitude of the parameters, particularly the origin of strong on-site repulsion $U_i$, we carry out the following simple estimation. For the phonon-disorder interaction Hamiltonian $H_I$, the first-order self-energy for the $i^{\text{th}}$ SL layer can be computed as
\begin{equation}
    \Sigma^1_i(\omega) =  g_i^2 \Bigg[\frac{n_B(\varepsilon_{d,i})+n_B(\varepsilon_i)}{\omega-\varepsilon_{i}+\varepsilon_{d,i}} + 
    \frac{n_B(\varepsilon_{d,i})-n_B(\varepsilon_i)+1}{\omega-\varepsilon_{i}-\varepsilon_{d,i}} \Bigg] 
    \label{eq:greensfunction}
\end{equation}
where $n_B$ is the Bose-Einstein distribution function. If we assume that the energy of disorder at a given SL layer is much smaller than the energy of SL, $\varepsilon_{d,i} \ll \varepsilon_{i}$, Eq. \eqref{eq:greensfunction} can be simplified as 
\begin{equation}
    \Sigma^1_i(\omega) \sim 2 g_i^2 \frac{n_B(\varepsilon_{d,i})}{\omega-\varepsilon_{i}}  
    \label{eq:greensfunction2}
\end{equation}
 The corresponding relaxation time can be computed as the imaginary part of Eq. \eqref{eq:greensfunction2}. The relaxation time of the $i^{\text{th}}$ SL layer $\tau_i$ can be written as 
\begin{equation}
    \frac{1}{\tau_i} = 4\pi g_i^2 n_B(\varepsilon_{d,i}) \nu(\epsilon_i)
    \label{eq:relaxationtime1}
\end{equation}
\noindent where $\nu(\varepsilon_{i})$ is the density-of-state in the $i^{\text{th}}$ SL. Furthermore, assuming Matthiessen's rule, that the total relaxation rate $1/\tau$ is the sum for each layer, we have $1/{\tau} \sim N/{\tau_i}$ if assuming other quantities are layer-independent or drawn from an independent random distribution. On the other hand, for phonon scattering with finite-sized nanodot disorder, in the regime $qR>1$, where $q$ is the phonon wavevector and $R$ is the disorder radius, it is known that geometrical scattering will dominate \cite{faleev2008}, such that
\begin{equation}
    \frac{1}{\tau_0} = 2\pi \eta v_s  R^2
    \label{eq:relaxationtime2}
\end{equation}
where $v_s$ is the sound velocity. Since Eqs. \eqref{eq:relaxationtime1} and \eqref{eq:relaxationtime2} are derived differently, but represent the same context for disorder scattering, we can assume that $1/{\tau} \propto 1/{\tau_0}$ without loss of generality. As a result, we have 
\begin{equation}
    g_i^2 \propto \frac{\eta v_s R^2}{2 N n_B(\varepsilon_{d,i}) \nu(\epsilon_i)}.
    \label{eq:gi2}
\end{equation}
For the non-interacting disorder part $H_d$, we adopt a total energy argument: the added disorder creates the energy that is proportional to the disorder concentration $\eta$, while the phonon energy is inversely proportional to the square root of the mass, $\varepsilon\propto\sqrt{1/m}$. As a result, the total disorder energy $\varepsilon_{d,i}$ and the SL phonon energy $\varepsilon_i$ are linked as  
\begin{equation}
    \varepsilon_{d,i} \propto \eta\sqrt{\frac{m_{\text{eff}}}{m_{\text{Er}}}}\varepsilon_i
    \label{eq:defectenergytophononenergy}
\end{equation}
where if we focus on phonon modes of Er, Ga and Al atoms, $m_{\text{eff}}=m_{\text{Ga}}m_{\text{Al}}/(m_{\text{Ga}}+m_{\text{Al}})$ is the effective mass of SL. Finally, using Eqs. \eqref{eq:gi2} and \eqref{eq:defectenergytophononenergy}, we have the on-site repulsion
\begin{equation}
    U_i = \frac{g_i^2}{\varepsilon_{d,i}} \propto \frac{\nu_s R^2}{Nn_B(\varepsilon_{d,i})\nu(\varepsilon_i)\varepsilon_i}\sqrt{\frac{m_{\text{Er}}}{m_{\text{eff}}}}.
    \label{eq:onsiterepulsion}
\end{equation}
From Eq. \eqref{eq:onsiterepulsion}, we observe a number of features that are directly associated with experimental observations, as discussed in main text. 

\subsection*{Data analysis of GI-IXS measurements}
\begin{figure}[ht]
	\centering
	\includegraphics[width=0.5\linewidth]{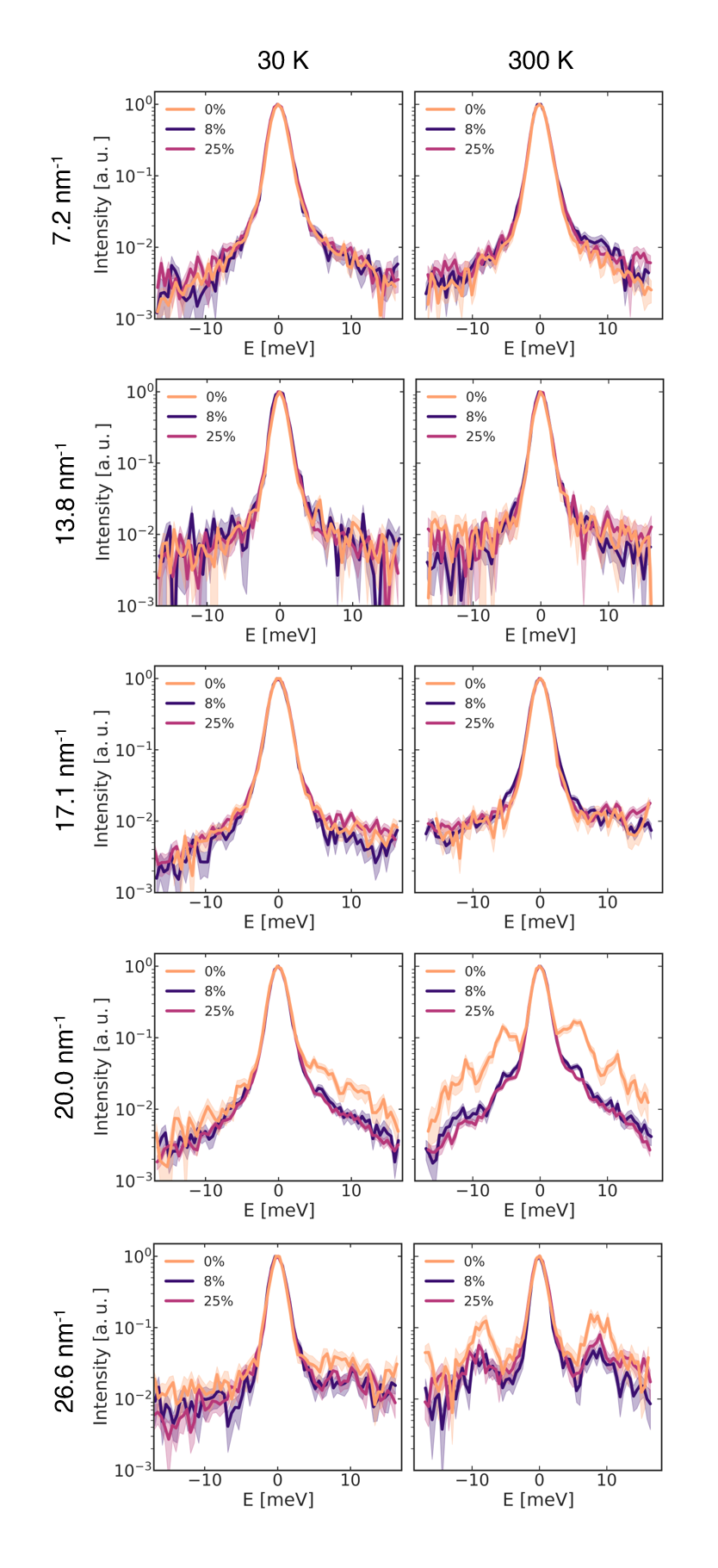}
	\caption{GI-IXS scans at constant-$q$ for samples with 0\%, 8\% and 25\% ErAs at 30 K and 300 K. The intensity is normalized and the pale regions indicate one standard deviation.}
	\label{sup:figure1}
\end{figure}

\noindent Constant-$q$ scans were performed, at 30 K and 300 K, in the GI-IXS configuration on the GaAs/AlAs samples containing 0\% (Ref), 8\% and 25\% ErAs nanodot disorder (percentage indicates the areal density) as shown in Figs. \eqref{fig:2} and \eqref{sup:figure1}. Within a constant-$q$ scan, the intensity of counts was measured as a function of incident x-ray energy with an energy spacing of 0.5 meV, reaching the limiting resolution of the instrument. Each intensity spectrum was normalized to the maximum intensity located at the Bragg peak and data points that fell outside of a symmetric interval about zero energy were not used. In the analysis, statistical error is taken to be the square root of the number of counts from Poisson statistics, corresponding to one standard deviation.

\begin{figure}[ht]
	\centering
	\includegraphics[width=0.5\linewidth]{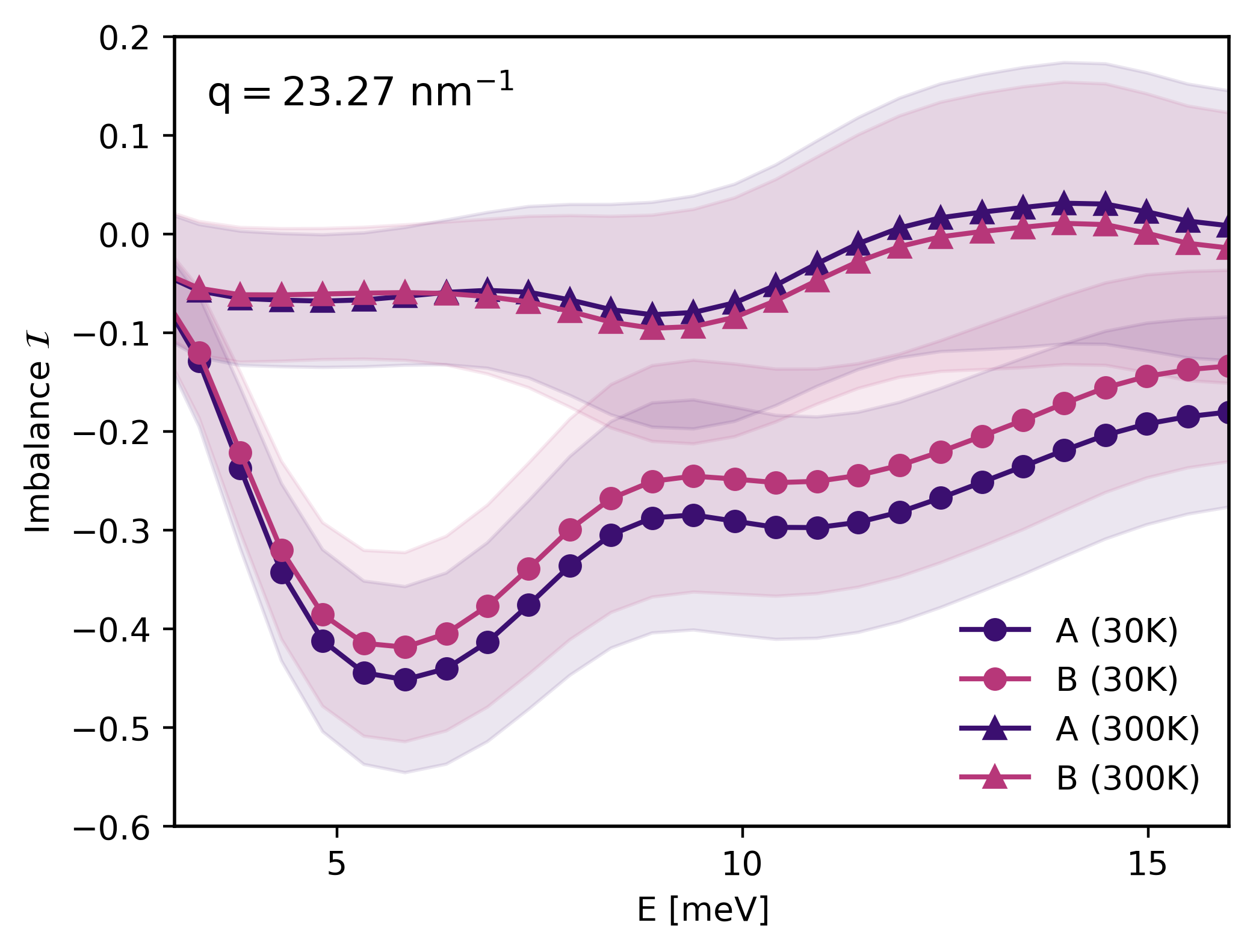}
	\caption{Imbalance parameter $\mathcal{I}$ as a function of incident x-ray energy for the GI-IXS scans at $q = 23.27$ nm$^{-1}$. Shaded pale regions indicate one standard deviation, calculated from the statistical error of the intensity counts. The MBL region, seen as a dip near 5 meV for disordered samples at 30 K (circular markers), displays an imbalance parameter that is four standard deviations away from zero. This is in contrast to the case at 300 K (triangular markers) which locate near zero within error.}
	\label{sup:figure2}
\end{figure}

To calculate Eq. \eqref{eq:pop_imbalance} for Fig. \ref{fig:3}, we begin by calculating $R_{\text{Disorder}}$ corresponding to either 8\% or 25\% ErAs nanodot disordered samples and $R_{\text{Ref}}$, the sample without ErAs disorder. $R = I(\text{Stokes})/I(\text{anti-Stokes})$ is the ratio between the intensities of Stokes scattering (positive energies) and anti-Stokes scattering (negative energies). This ratio was taken while linearly interpolating the intensity between the adjacent data points. It was noticed that, at the highest energies of each run, the intensity would sometimes drop to an extremely low value which would make the ratio $R$ drastically diverge. As such, a constant value of $1\times 10^{-5}$ was added to the intensity of each data point. The addition of this constant does not significantly alter the results as the value corresponds to two orders of magnitude less than the overall measured background and at most, it would result in a slight under-estimation of $R$ at the highest measured energies, which does not impact the major conclusions drawn from this study. A line cut of Fig. \ref{fig:3} at $q = 23.3$ nm$^{-1}$ at 8\% and 25\% ErAs disorder is shown in Fig. \ref{sup:figure2} with an indication of error for $\mathcal{I}$. The error on $\mathcal{I}$ is calculated by properly distributing the error of the intensity counts, which follow Poisson statistics. Notably, the MBL phase region, whereby $\mathcal{I} \sim 0.5$, lies four standard deviations away from the trivial $\mathcal{I} = 0$ phase at this value of $q$.

As mentioned in the main text, the calculation of $R$ (and therefore $\mathcal{I}$) shown in Fig. \ref{fig:3} was performed without removal of the elastic peak in order to rule out possible artifacts that may arise from such a subtraction. In addition, Fig. \ref{fig:3} does not consider the effect of the resolution functions from the analyzers that were used in the measurement scans. Indeed, a comparison of the data with the resolution profile is important to properly evaluate the balance factor of the spectra and the suppression of the acoustic phonon population necessary for MBL (Eq. \eqref{eq:Ui}). Due to instrumental constraints, the resolution functions of the analyzers were measured using poly(methyl methacrylate) (PMMA) in the transmission geometry. However, from previous knowledge of small-angle neutron scattering (SANS) and the respective grazing incidence version (GI-SANS) \cite{muller-buschbaum2013}, the resolution functions between transmission geometry and reflection geometry (which was used for the GaAs/AlAs samples) may differ significantly and transformation between the two is difficult. As such, deconvolution of the spectra is not ideally feasible due to the geometry of the resolution function and the relatively low signal-to-noise ratio from a grazing-incidence measurement. As such, we opt to analyze and present the original data in its raw form (Fig. \ref{fig:3}) and provide the population imbalance with an error bar (Fig. \ref{sup:figure2}) which includes errors from the resolution function (such as any asymmetry). Possible asymmetries of the resolution function with respect to zero energy were examined and concluded to not contribute to the observed imbalance in $\mathcal{I}$. Nevertheless, we present below a more-thorough analysis which considers these aspects yet does not take away from the results of the main text.

\begin{figure}[ht]
	\centering
	\includegraphics[width=\linewidth]{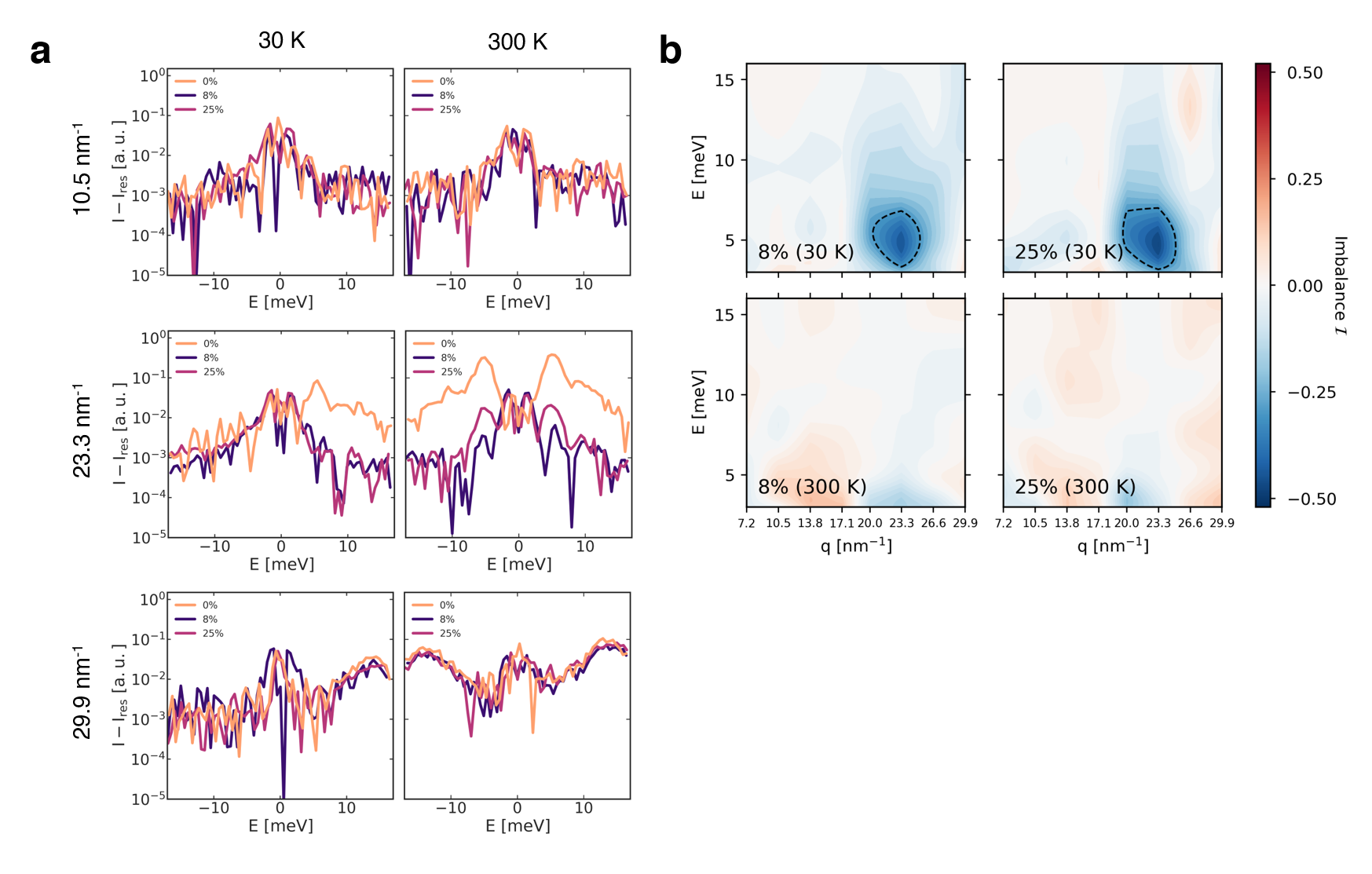}
	\caption{\textbf{a}, Intensity counts for the GI-IXS scans at constant-$q$ after subtraction with the resolution function of the analyzer (normalized to the maximum intensity) for samples with 0\%, 8\% and 25\% ErAs at 30 K and 300 K. \textbf{b}, The colormap of the phonon population imbalance $\mathcal{I}$ at two temperatures (30 K and 300 K) is plotted for ErAs-disordered SLs (8\%, left column, and 25\%, right column) as defined in Eq. \eqref{eq:pop_imbalance} using the intensities obtained after the subtraction with the analyzer resolution function. Reminiscent of Fig. \ref{fig:3}, the imbalance is observed at low temperature in similar energy and wavevector windows. One notices that removal of the resolution function decreases the value of the imbalance at larger energies (above 10 meV), where the intensity counts are lower. The dashed lines denote the contour for $|\mathcal{I}|=0.3$.}
    \label{sup:figure3}
\end{figure}

\begin{figure}[ht]
	\centering
	\includegraphics[width=\linewidth]{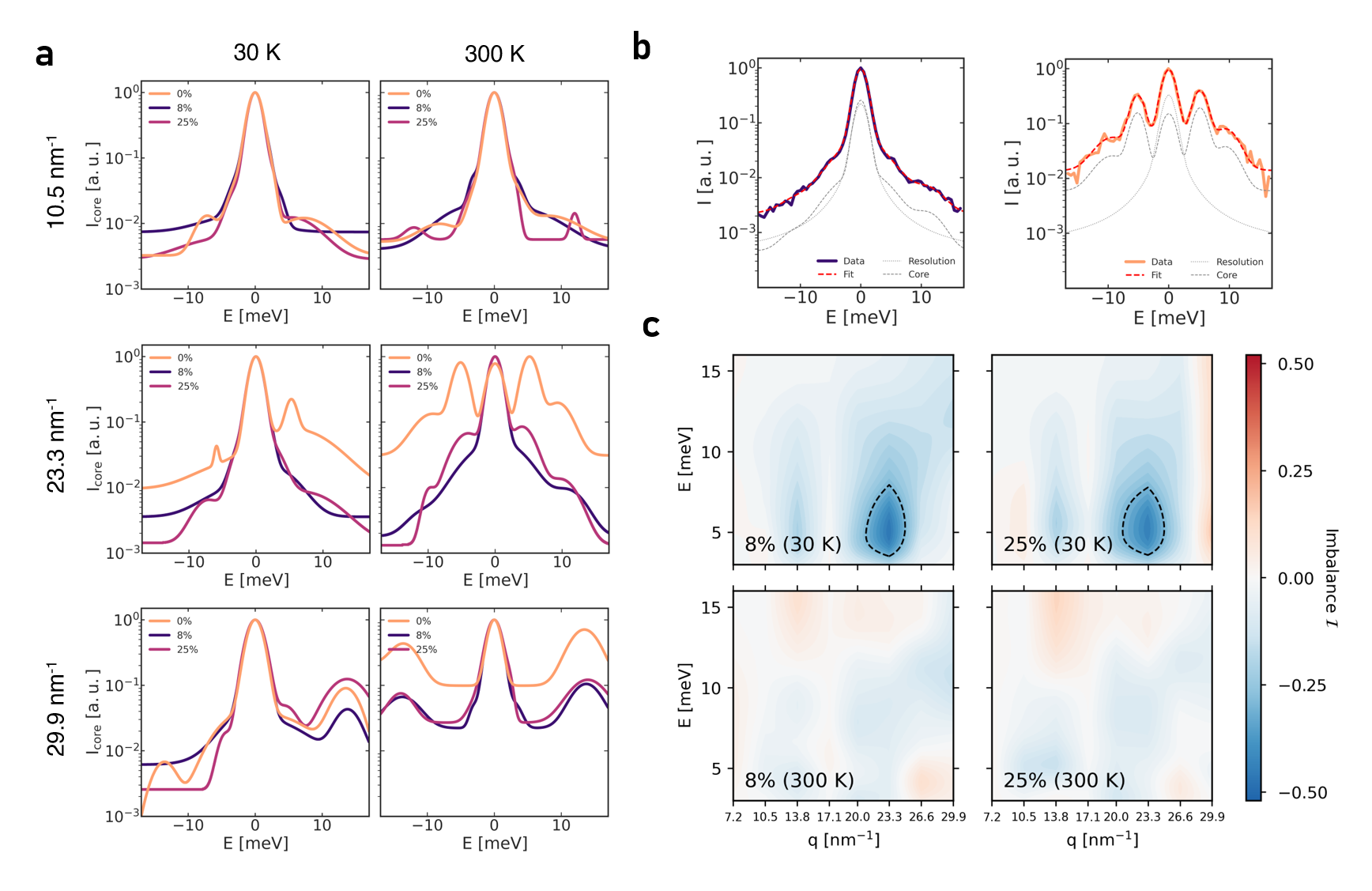}
	\caption{\textbf{a}, Intensity counts for constant-$q$ scans of the extracted core function representing the phonon modes after deconvolution with the resolution function (normalized to maximum intensity) for samples with 0\%, 8\% and 25\% ErAs at 30 K and 300 K. The core function corresponds to damped harmonic oscillator model (Lorentzian peaks) with a uniform background. \textbf{b}, Examples of deconvolution of the GI-IXS spectrum (data points in color) with the resolution peak (light-grey dotted line). The core function (grey dashed line) is extracted from the overall fit (red dashed line) with good agreement. \textbf{c}, The colormap of the phonon population imbalance $\mathcal{I}$ at two temperatures (30 K and 300 K) is plotted for ErAs-disordered SLs (8\%, left column, and 25\%, right column) as defined in Eq. \eqref{eq:pop_imbalance} using the extracted core function after deconvolution with the resolution function. A strong imbalance is seen at similar energy and wavevector intervals as Fig. \ref{fig:3} of the main text. The dashed lines denote the contour for $|\mathcal{I}|=0.3$.}
    \label{sup:figure9}
\end{figure}

In this supplementary note, we present a coarse examination of the data which considers the resolution functions of the analyzers used in the experiment. The following two methods are performed to demonstrate that similar results are obtained, independent of data handling procedures. (1) A direct subtraction of the resolution function from the data (with normalization taken into account) shown in Fig. \ref{sup:figure3}. (2) A crude fit to the data using a convolution of an improvised model core function, that regards discernible peaks as Lorentzian peaks (damped harmonic oscillator model), with the analyzer resolution function (treated as a pseudo-Voigt function) shown in Fig. \ref{sup:figure9}. We show that even with such considerations the conclusions of the main text concerning the data remain valid: the population imbalance is largest at 30 K and $q = 23.3$ nm$^{-1}$ within an energy interval of 4-10 meV (with the strongest deviation between 5-6 meV). This supports our observations from raw data on imbalance, and supports our interpretation of this population imbalance result as an indication of a many-body-localized phase of phonons.

\subsection*{Phonon dispersion of GaAs/AlAs/ErAs superlattices}
The phonon dispersion along the path $\mathbf{q}=(0, 0, q_z)$ is computed and the $z$ direction is normal to interfaces. The period length for superlattice structure is $\sim$ 3 nm. Due to the supercell of superlattice structure, the phonon dispersion is folded into the Brillouin zone of the supercell, making the analysis of phonon spectrum difficult. To better study the relationship between phonon frequency and momentum, we compute the unfolded phonon dispersion within the Brillouin zone of perfect bulk [001] GaAs. The algorithm of unfolding the phonon dispersion can be found in Ref. \cite{allen2013}. The width of the dispersion indicates the amount of weight at given unfolded wave vector. The source code can be found in Ref. \cite{song2020}. The interatomic force constant required for dispersion calculation is obtained from density function perturbation theory (DFPT) \cite{baroni2001} for a 4-atom [001] GaAs unit cell. We assume the force constants for GaAs, AlAs and ErAs are the same such that the mass difference for different atoms causes differences in dynamical matrix as well as the phonon dispersion. A 6$\times$6$\times$6 $q$ point mesh is used for DFPT calculation and 8$\times$8$\times$8 $k$ point mesh is used for DFT calculation, implemented in the \textlcsc{Quantum Espresso} package \cite{giannozzi2009}. We have used norm-conserving pseudopotentials with local density approximation (LDA) for exchange-correlation energy functional and a cut off energy of 72 Rydberg in the DFT calculation.

\begin{figure}[ht]
	\centering
	\includegraphics[width=\linewidth]{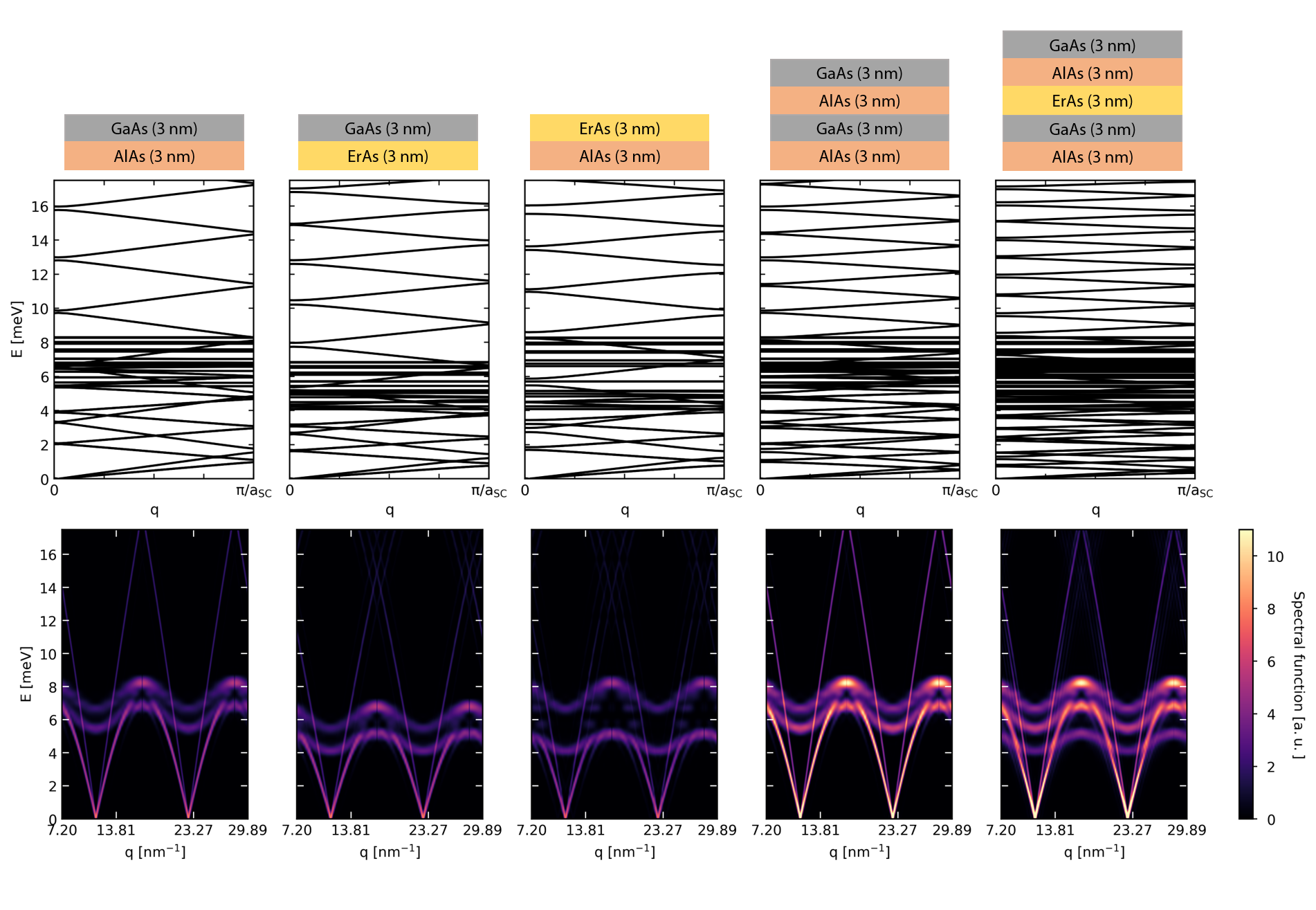}
	\caption{Theoretical phonon dispersions of superlattices containing 3 nm layers of GaAs, AlAs and ErAs. From left to right, schematics of the following scenarios are shown at the top: i) GaAs/AlAs, ii) GaAs/ErAs, iii) ErAs/AlAs, iv) GaAs/AlAs/GaAs/AlAs, and v) GaAs/AlAs/ErAs/GaAs/AlAs. In the middle row, the corresponding calculated, unfolded phonon dispersion is plotted as a function of the wavevector $q$ within the mini-Brillouin zone of the superlattice from $q = 0$ up to $q = \pi/a_{SC}$ where $a_{SC}$ is taken to be $\sim$5.55 \AA. In the bottom row, the theoretical phonon dispersion is plotted as a function of wavevector for $q$-values achieved in the GI-IXS experiment. The intensity of the colormap, in arbitrary units, indicates the value of the calculated spectral function.}
    \label{sup:figure8}
\end{figure}

\begin{figure}[ht]
	\centering
	\includegraphics[width=0.75\linewidth]{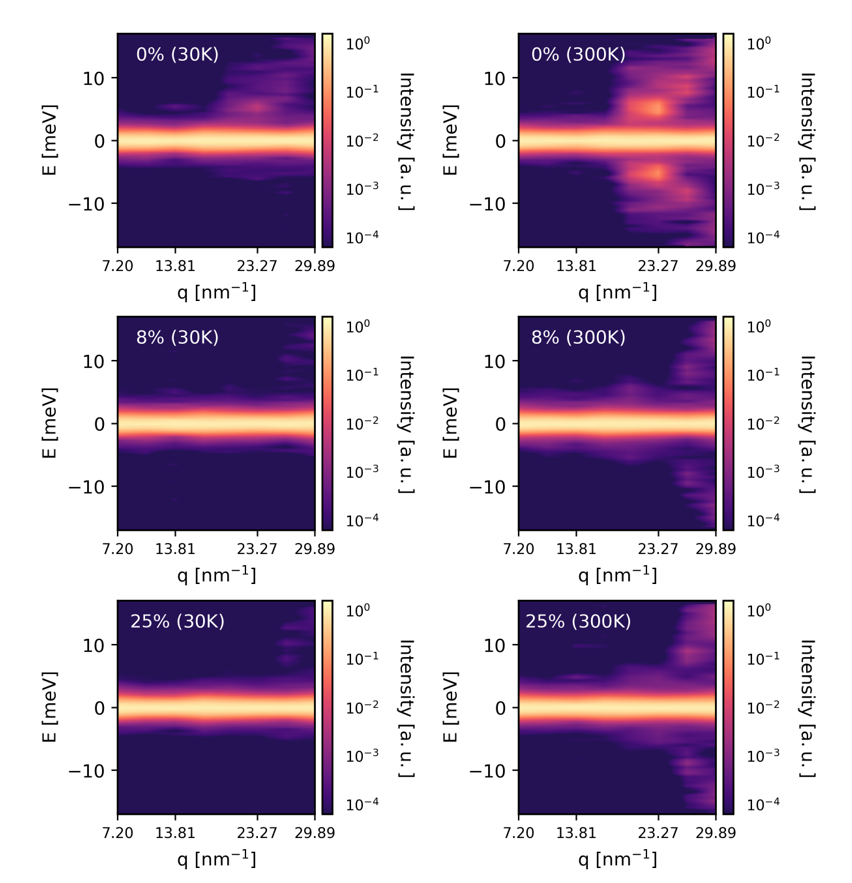}
	\caption{Phonon dispersion relations obtained from GI-IXS spectra at 30 K and 300 K for samples with 0\%, 8\% and 25\% ErAs. The colormap is representative of the normalized intensity in log-scale.}
    \label{sup:figure4}
\end{figure}

The simulation of ErAs nano-dot disorder in GaAs/AlAs superlattices is computationally intractable due to the randomly displaced positions and the necessity of tracking a large number of degrees of freedom. Thus, a complete phonon dispersion calculation of this scenario is difficult to achieve. Instead, we compute phonon dispersions of lattices composed of a finite number of layers of GaAs, AlAs or ErAs (of thickness 3 nm) and contrast the differences between the different materials and total thicknesses. These theoretical phonon dispersions using the procedure described above are shown in Fig. \ref{sup:figure8}, for the unfolded phonon dispersion within the mini-Brillouin zone and within the wavevector range accessed in the GI-IXS experiment. Despite differences in the material type and the number of layers, there is a broad range demonstrating high intensities of the spectral function for the phonon dispersion between 5-10 meV. This energy interval corresponds to the main bulk of phonon energies that are experimentally obtained, as shown in the measured phonon dispersion of Fig. \ref{sup:figure4} for different ErAs disorder levels and different temperatures.

\subsection*{Effect of GaAs substrate}
\begin{figure}[ht]
	\centering
	\includegraphics[width=\linewidth]{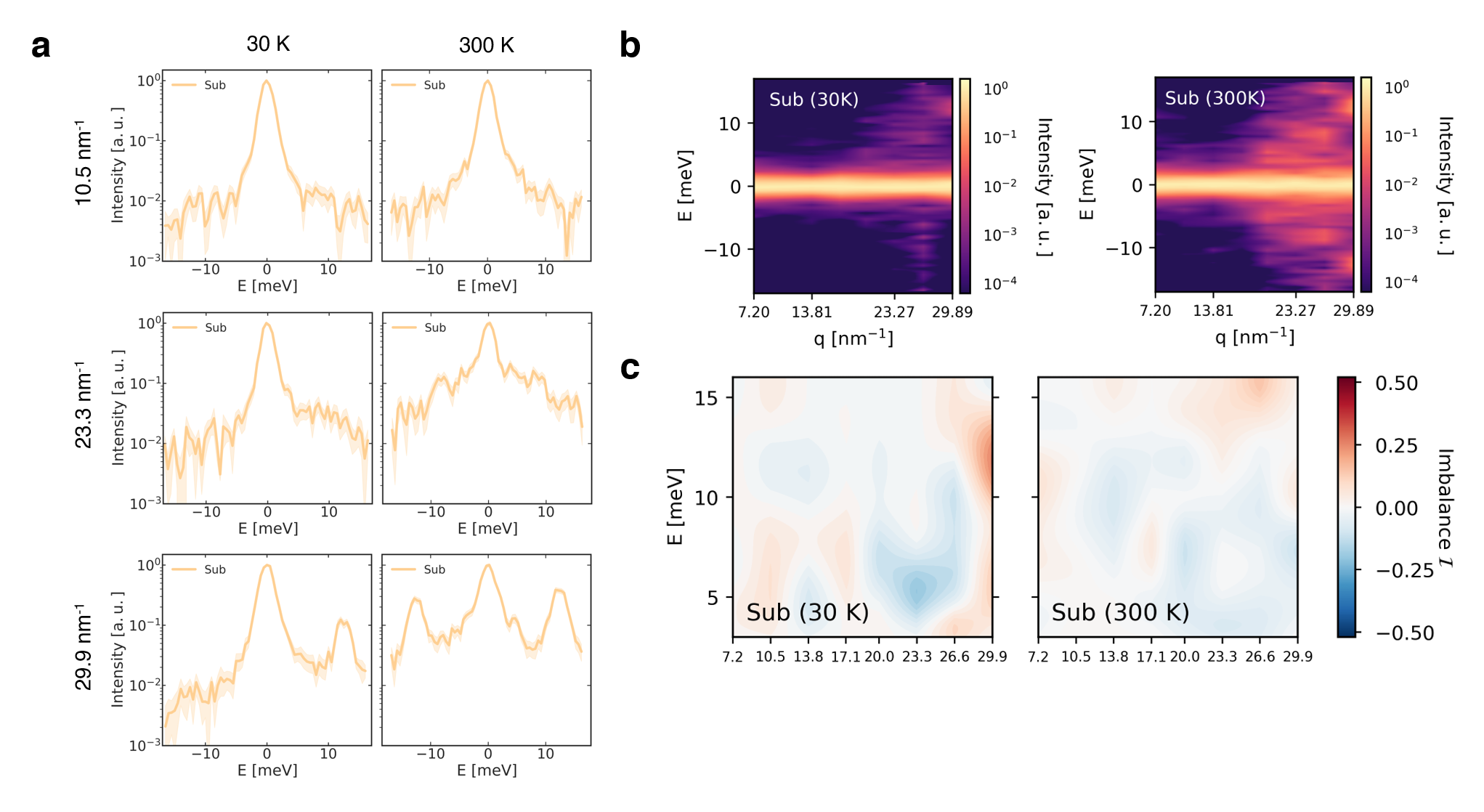}
	\caption{\textbf{a}, GI-IXS scans at constant-$q$ for the GaAs substrate at 30 K and 300 K. The intensity is normalized and the pale regions indicate one standard deviation. \textbf{b}, Corresponding phonon dispersion relations plotted as a colormap in energy and wavevector space. \textbf{c}, Phonon population imbalance $\mathcal{I}$ for the GaAs substrate compared with the Ref SL using \eqref{eq:pop_imbalance}. The large imbalance signature seen in disordered SLs (as in Fig. \ref{fig:3}) is not seen for the GaAs substrate, thereby excluding any involvement of the substrate on the observed imbalance.}
    \label{sup:figure7}
\end{figure}

Identical GI-IXS measurements were taken on the GaAs substrate without the GaAs and AlAs SL structure on top. These are shown as individual constant-$q$ scans and as a phonon dispersion colormap in Figs. \hyperref[sup:figure7]{S7a}-\hyperref[sup:figure7]{S7b}, respectively. One should note, as mentioned in the previous section, that the phonon dispersion for solely the GaAs substrate is distinct from that of the SLs. We apply the same data analysis as was executed for the disordered SLs in comparison to the Ref SL to discern whether there are important substrate contributions to the observed population imbalance in Fig. \ref{fig:3}. A plot of the population imbalance $\mathcal{I}$ is shown in Fig. \hyperref[sup:figure7]{S7c}. The maximum imbalance measured for the GaAs substrate is only $\sim$ 0.20, significantly lower than the 0.50 value measured for disordered SLs. Moreover, the distinctive energy and wavevector intervals wherein the population imbalance decreases significantly (which was the case for both disordered SLs (8\% and 25\%)) are not observed. Accordingly, these observations from GI-IXS on purely the GaAs substrate support the idea that it does not play a role in the abnormal population imbalance seen for the disordered SLs.

\subsection*{Raman scattering}
We measured Raman spectra of GaAs/AlAs SLs with variations on ErAs nano-dot doping levels (0\%, 8\% and 25\% by areal density) and on the temperature. These were measured in a custom-built micro-Raman setup. The SLs were excited with a continuous wave diode-pumped solid-state laser (Excelsior, Spectra Physics, 532 nm, 100 mW) through an upright microscope using a 50x distance objective with a numeric aperture of 0.5. The typical incident laser power on the sample was 500 $\mu$W. The scattered Raman light was analyzed by a spectrometer (Spectra Pro 2300i, Acton, $f = 0.3$ m) that was coupled to the microscope and equipped with a 1800 grooves/mm grating along with a CCD camera (Pixis 256BR, Princeton Instruments). The Raman spectra at low temperatures were measured in a liquid He-cryostat (MicrostatHiResII, Oxford Instruments) with a temperature controller (MercuryiT, Oxford Instruments) that allowed precise temperature control. The cryostat was mounted on a motorized XY microscope stage and evacuated to $7\times10^{-7}$ mbar prior to cool down.

\begin{figure}[ht]
	\centering
	\includegraphics[width=0.75\linewidth]{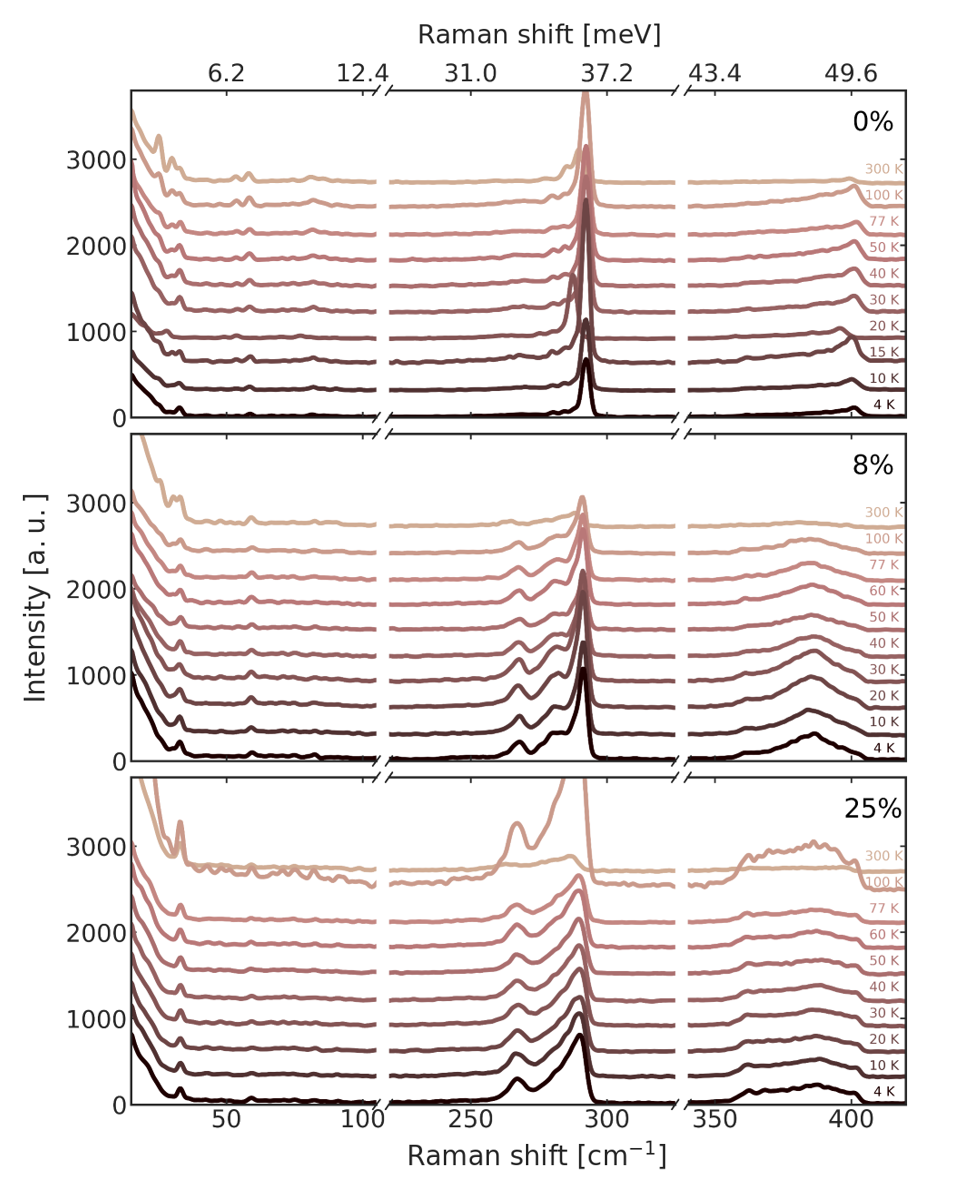}
	\caption{Raman spectra at different temperatures ranging from 4 K to 300 K for GaAs/AlAs SLs with 0\% (top), 8\% (middle) and 25\% (bottom) ErAs nano-dot order by areal density. Spectra are offset by 300 (arbitrary units) for each temperature for visual clarity.}
    \label{sup:figure5}
\end{figure}

We present the data in Fig. \ref{sup:figure5} for the different 300-period GaAs/AlAs SLs at numerous temperatures. The results of the Raman scattering are consistent with previous thermal conductivity measurements using time-domain thermoreflectance \cite{luckyanova2018}. The peak at approximately 28 cm$^{-1}$ corresponds to an air peak artifact. At first glance, one notices the existence of phonons in all three types of systems through peaks at approximately 50-60 cm$^{-1}$ (corresponding to 6.2-7.4 meV). This is namely the energy range where the phonon population imbalance of Fig. \ref{fig:3} is the largest. Furthermore, at larger Raman shifts, outside the energy range of the x-rays used in the GI-IXS experiment, the introduction of ErAs disorder (either 8\% and 25\%) gives rise to an additional Raman peak near 266 cm$^{-1}$ and a broad peak feature at 350-400 cm$^{-1}$.  

\begin{figure}[ht]
	\centering
	\includegraphics[width=0.75\linewidth]{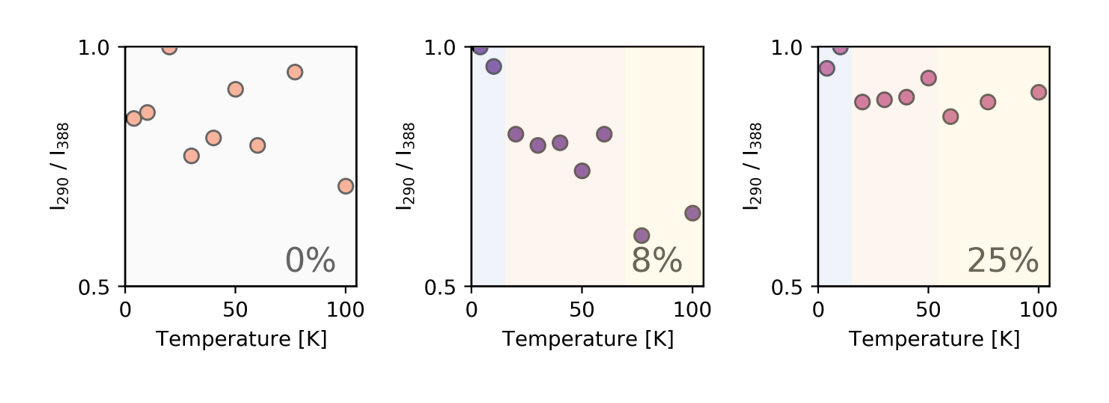}
	\caption{Ratio of the intensity of the Raman peak at 290 cm$^{-1}$ with that at 388 cm$^{-1}$ normalized to the maximum intensity ratio at 0\%, 8\% and 25\% ErAs-nanodot disorder. For the disordered SLs, pale-colored regions serve as a visual indication of the possible phases revealed through these intensity ratios within certain temperature intervals.}
    \label{sup:figure6}
\end{figure}

For these 300-period SLs containing ErAs disorder, we observe a slight indication of a distinct phase upon comparing peak intensities with phonon frequencies of approximately to 290 cm$^{-1}$ and 388 cm$^{-1}$ as shown in Fig. \ref{sup:figure6}. This phase occurs within a temperature interval of 20 K to 50-70 K which comprises the 30 K GI-IXS measurements that were shown in the main text. The mechanism behind the dramatic decrease in the intensity ratio for the 8\% ErAs-disordered sample relative to the 25\% remains elusive. The occurrence of this phase seen as plateaus in the intensity ratio does not appear for the Ref sample for which there is no discernible structure.

\clearpage

\clearpage

\clearpage

\clearpage

\clearpage

\clearpage

\clearpage

\clearpage

\clearpage

\end{document}